\documentstyle[12pt]{article}
\begin{document}
\baselineskip=21pt
\begin{titlepage}
\rightline{Alberta-Thy-46-93}
\rightline{SEPTEMBER 1993}
\vskip .5in
\begin{center}
{\large{\bf TOPOLOGICAL MASS GENERATION IN THREE-DIMENSIONAL STRING
THEORY}}
\end{center}
\vskip .1in
\begin{center}

{\bf Nemanja Kaloper}
\vskip.2in
{\it Theoretical Physics Institute}

{\it Department of Physics, University of Alberta}

{\it  Edmonton, Alberta, Canada T6G 2J1}
\vskip.2in
\vskip.5cm
\end{center}
\centerline{ {\bf Abstract} }
\baselineskip=18pt

\noindent
The effective action
of string theory in three dimensions is investigated,
incorporating the Lorentz and gauge Chern-Simons
terms in the definition of the Kalb-Ramond axion field strength. Since
in three dimensions any three-form is trivial,
the action can be reformulated by properly integrating the axion out.
The circumstances under which it can be recast in form
of topologically massive gravity coupled to a topologically massive
gauge theory are pointed out. Finally, the strong coupling limit of the
resulting action is inspected, with the focus on the roles played by the
axion and dilaton fields.

\vskip1cm
\centerline{\it Submitted to Phys. Lett. {\bf B}}
\end{titlepage}

\baselineskip=21pt

\newpage
{\newcommand{\la}{\mbox{\raisebox{-.6ex}{$\stackrel{<}{\sim}$}}}
{\newcommand{\ga}{\mbox{\raisebox{-.6ex}{$\stackrel{>}{\sim}$}}}

The construction of a consistent quantum theory of gravity, and its merger
with the theories of other interactions in Nature, has remained an
outstanding problem of modern physics for a long time. The only candidate
to lay a claim to achieving such a unification so far is string theory. Its
development as a theory formulated in curved target manifolds in recent
years has provided us with new insights in the structure of gravity.
String theory can be consistently formulated in a variety of target
space dimensions. Although we know that Nature picks the number of
dimensions of the space-time to be four, the theories in dimensions other
than four are also useful, because they provide us with a
theoretical laboratory to study string theory and gravity, in the absence
of an adequate experimental one. Here some aspects of the background field
formulation of three-dimensional (3D) string theory will be addressed.
Motivation for the study of a 3D model comes from several
directions. It is
well known \cite{WDim} that  such a model can be obtained
from 10D superstring theory by dimensional reduction. Hence its
properties might bear on the structure of higher-dimensional theories too.
Furthermore, it can also be realized as a factor of a consistent
direct product string theory, where the other
factor is some internal conformal field theory.
The very simple mathematical structure of 3D gravity,
when compared to higher-dimensional
formulations, renders it a convenient toy-model
to investigate certain properties of gravity, elusive in more complicated
cases. This is well illustrated by a rapidly growing interest in
various guises of 3D theories of gravity in the past decade or so
\cite{DJT}-\cite{sols}.

The purpose of this letter is to investigate the relationship between
$O(\alpha')$ 3D string theory and  the
topologically massive gravity (TMG) of
Deser, Jackiw and Templeton \cite{DJT}.
Specifically, I will show that a TMG look-alike represents
a special sector of 3D string theory, when the dilaton field is constant.
A related argument was offered by Nishino and
Nishino and Gates \cite{NGN}, but in their
construction it was necessary to resort to a dual formulation of the
Kalb-Ramond axion in 10D superstring theory and then dimensionally
reduce the model down to 3D. I will show here that it is possible to establish
such a relationship within the 3D theory itself, without resorting
to higher dimensions and dimensional reduction. It will be essential
for the argument
to note that in three dimensions, the Kalb-Ramond field, being a three-form,
has no dynamics, and is basically a constant topological condensate,
proportional to the volume element. The implementation of this observation
at the level of the action, with the help of a Lagrange multiplier method
analogous to that of the standard duality transformations in higher
dimensions, establishes the above mentioned relationship. In the end,
both gravitational and gauge fields become topologically
massive. I will analyze certain properties of the resulting action,
focusing on the role the axion and dilaton fields play in giving rise
to the gravity and gauge topological mass terms. These conspire in the
model in a special way, resulting in the axion condensate
giving rise to the explicit topological mass terms,
and the dilaton v.e.v. manifestly breaking conformal invariance.
The analysis indicates that
in the limit of strong string coupling $<{\bf \Phi}> \rightarrow \infty$,
the conformal symmetry can be restored, and the theory returns to
the symmetric phase, described to the lowest order
as a direct sum of Chern-Simons gravity and
a Chern-Simons gauge theory. This, in light of the results of
\cite{WHW}, can be understood as the
$SO(3,2) \times G$ Chern-Simons gauge theory.

The starting point of the investigation to be carried out here is the
world-sheet action for the background field formulation of 3D string theory.
For simplicity's sake,
I choose to work in Planck units, setting
$\kappa^2 = 1$, on the tangent space spanned by the dreibein
$\Bigl\{{\bf e^{a}}\Bigr\}$. Forms will
be represented with bold-face symbols, whereas
all other objects are tensor components. The wedge symbols are suppressed for
simplicity, and are asumed whenever a
product of two forms is written. The conventions are
${\bf \alpha} \wedge {\bf \beta} =
\frac{(p+q)!}{p!q!} Alt ({\bf \alpha} \otimes {\bf \beta})$.
To order $O(\alpha')$, the terms containing {\it effectively} up to three
derivatives are  given  by
\begin{equation}
S~=~-\frac{1}{2} \int  e^{-{\bf \Phi}}
\Bigl\{\epsilon_{abc}{\bf e}^a {\bf R}^{bc} + d{\bf \Phi}~{^{*}d}{\bf \Phi}
- 2 {\bf H}~{^{*}{\bf H}}
+{\alpha' \over 8} Tr {\bf F}~{^{*}{\bf F}}
+ \frac{\Lambda(\Phi)}{3}
\epsilon_{abc}{\bf e}^a{\bf e}^b{\bf e}^c~\Bigr\}
\label{act}
\end{equation}

\noindent  The normalization conventions are adopted from
\cite{GS}. The gauge field strength of the
1-form gauge potential ${\bf A}$ is $~{\bf F} = d{\bf A} + {\bf A}^2~$,
$~{\bf H} = d{\bf B}
+ (\alpha'/16) \Bigl( {\bf \Omega}_L - {\bf \Omega}_{YM} \Bigr)~$
is the field strength associated with the Kalb-Ramond 2-form
field $~{\bf B}~$ and the 0-form $~{\bf \Phi}~$
is the dilaton field. The Chern-Simons forms
\begin{eqnarray}
&&{\bf \Omega}_L{} = Tr\bigl( {\bf \omega} d{\bf \omega}
+ \frac{2}{3} {\bf \omega}^3 \bigr)
 \nonumber \\
&&{\bf \Omega}_{YM}{} = Tr\bigl( {\bf A}d{\bf A} + \frac{2}{3} {\bf A}^3 \bigr)
\label{csa}
\end{eqnarray}

\noindent appear
in the definition of the axion field strength due to the Green-Schwarz anomaly
cancellation mechanism, and can be understood as a model-independent
residue after dimensional reduction from ten-dimensional superstring theory.
The gauge group generators $X^A$ are normalized according to
$~Tr X^A X^B = - 2 \delta^{AB}~$. The dilaton self-interaction potential
$\Lambda(\Phi)$ has been included keeping in mind the possibility that
it may arise from non-perturbative effects, as well as in the presence
of the stringy cosmological constant. Generically,
$~\Lambda(\Phi)=\Lambda_0 + W(\Phi)~$, with $~\Lambda_0~$ being the
cosmological constant which may arise as the difference of the
internal theory central charge and the total central charge for a
conformally invariant theory, and
$~W(\Phi)~$ being the dilaton self-interaction of non-perturbative
origin \cite{DRSW}. Other terms will be ignored, since they
can be viewed as the corrections to the dynamics described
by (\ref{act}). Moreover, some of these will vanish identically in three
dimensions (e.g. Gauss-Bonnet), some can be reduced to the others, and some
removed altogether with the help of field redefinitions.

An indication that there exists a connection between the theory described by
(\ref{act}) and TMG can be obtained by inspecting the equations of motion
derived from (\ref{act}). They are
\begin{eqnarray}
e^{-\Phi}G_{ab} - \frac{\alpha'}{4}
\nabla_c\Bigl(e^{-\Phi}H_{de(a}R^{c}{}_{b)}{}^{de} \Bigr)=e^{-\Phi}
\Bigl(H_{acd}H_{b}{}^{cd}&-&\frac{1}{6} \eta_{ab} H^2
- \frac{\alpha'}{8} ~Tr F_{ac}F_{b}{}^{c} + \frac{\alpha'}{32} ~
\eta_{ab}~ Tr F^2
\nonumber\\
+\Lambda({\bf \Phi}) \eta_{ab}&-&\frac{1}{2} \eta_{ab}
\bigl(\nabla \Phi\bigr)^2
- \nabla_a \nabla_b \Phi + \eta_{ab} \nabla^2 \Phi \Bigr) \nonumber\\
e^{-{\bf \Phi}}\Bigl\{d{^{*}d} {\bf \Phi}
- \frac{1}{2} d{\bf \Phi} {^{*}d}{\bf \Phi} + \frac{1}{2}
\Bigl( \epsilon_{abc}{\bf e}^a {\bf R}^{bc}&-&2 {\bf H}~{^{*}{\bf H}}
+ {\alpha' \over 8} Tr {\bf F}~{^{*}{\bf F}} \\
+ \frac{1}{3}\bigl(\Lambda({\bf \Phi})
&-&\frac{\partial \Lambda({\bf \Phi})}{\partial {\bf \Phi}} \bigr)
\epsilon_{abc}{\bf e}^a{\bf e}^b{\bf e}^c~\Bigr) \Bigr\} = 0 \nonumber  \\
D e^{-{\bf \Phi}}~{^{*}{\bf F}}
+ \frac{1}{2} e^{-{\bf \Phi}}~{^{*}{\bf H}}~{\bf F} = 0
{}~~~~~D{\bf F} = 0
{}~~~&&~~d e^{-{\bf \Phi}}~{^{*}{\bf H}} = 0
{}~~~~~d{\bf H} = 0 \nonumber
\label{EQS}
\end{eqnarray}

\noindent The operator $D$ above represents the gauge-covariant exterior
derivative $~D{\bf Z} = d{\bf Z} + \bigl[{\bf A},{\bf Z}\bigr]~$,
and the operator $\nabla_a$ is the standard covariant derivative.
Note that the equation representing the Bianchi
identity for the Kalb-Ramond field can be omitted.
In four- and higher-dimensional theories,
this identity, the result of the map of ${\bf H}$ by exterior derivative $d$,
reads $d{\bf H} = (\alpha'/16) \Bigl( Tr {\bf R}^2 - Tr {\bf F}^2 \Bigr)$.
However, in 3D both sides of this
equation, being  four-forms, are identically zero, and so the equation
$d{\bf H} = 0$ is satisfied {\it vacuously}. This is
a very important detail for establishing the connection
between (\ref{act}) and TMG. As a consequence,
$H_{abc} = f \epsilon_{abc}$ where $f$ is a function to be
determined such that the Euler-Lagrange equation for $~{\bf H}~$ is
solved. This yields $f = Q \exp(\Phi)$, where
$Q$ is a constant. In three dimensions
$~R_{abcd} = R_{ac}\eta_{bd}-R_{ad}\eta_{bc}+
R_{bd}\eta_{ac}-R_{bc}\eta_{ad} + \bigl(R/2\bigr)
\bigl(\eta_{ad}\eta_{bc}-\eta_{ac}\eta_{bd}\bigr)~$,
and so, using $H_{abc} =  Q \exp(\Phi) \epsilon_{abc}$ one finds that
$~\nabla_c\Bigl(e^{-\Phi}H_{de(a}R^{c}{}_{b)}{}^{de} \Bigr) = 2QC_{ab}~$,
where $~C_{ab}~$ is the Cotton tensor. The equations of motion (3)
can be rewritten as
\begin{eqnarray}
e^{-\Phi}G_{ab} - \frac{\alpha'}{2} Q C_{ab}&=&e^{-\Phi}
\Bigl(\frac{\alpha'}{32} ~\eta_{ab} ~Tr F^2
- \frac{\alpha'}{8} ~Tr F_{ac}F_{b}{}^{c}  \nonumber\\
&+&\bigl(\Lambda({\bf \Phi}) - Q^2 e^{2\Phi} \bigr) \eta_{ab}
- \frac{1}{2} \eta_{ab} \bigl(\nabla \Phi\bigr)^2
 - \nabla_a \nabla_b \Phi + \eta_{ab} \nabla^2 \Phi \Bigr)\nonumber\\
e^{-{\bf \Phi}}\Bigl\{d{^{*}d} {\bf \Phi}
- \frac{1}{2} d{\bf \Phi} {^{*}d}{\bf \Phi}&+&\frac{1}{2}
\Bigl(\epsilon_{abc}{\bf e}^a {\bf R}^{bc}
+ {\alpha' \over 8} Tr {\bf F}~{^{*}{\bf F}}\\
&+&\frac{1}{3}\bigl(Q^2 e^{2{\bf \Phi}} + \Lambda({\bf \Phi})
- \frac{\partial \Lambda({\bf \Phi})}{\partial {\bf \Phi}} \bigr)
\epsilon_{abc}{\bf e}^a{\bf e}^b{\bf e}^c ~\Bigr) \Bigr\} = 0 \nonumber\\
D e^{-{\bf \Phi}}~{^{*}{\bf F}}&-&2 Q~{\bf F} = 0
{}~~~~~~~~~~~D{\bf F} = 0 \nonumber
\label{Eqdu}
\end{eqnarray}

Obviously, if ${\bf \Phi} = {\rm const.}$, and the dilaton equation of
motion is ignored, the remaining equations are precisely those of
topologically massive gravity coupled to a topologically massive gauge
theory (in presence of a cosmological constant). That such a special case is
indeed possible can be seen from the dilaton equation of motion.
Namely, in the case when the total dilaton self-interaction
potential $V(\Phi)$ (including all the terms in
the second of equations (4)
except the dilaton kinetic terms) has a minimum, the dilaton decouples.
This is precisely the situation encountered in \cite{HW-K}, where the
contributions to the dilaton self-interaction arise from the cosmological
constant and the axion. As a corollary, gauge field
perturbations around the 3D stringy black hole background are topologically
massive.

The natural question to ask next is what is the action from which one can
derive the dynamics described by (4), and how is it related
to (\ref{act}). To answer this it is best to resort to dualizing
the Kalb-Ramond axion directly in the action.
Before one proceeds in this direction, it is instructive to recall how
this is performed in the more common case of four dimensions. There,
(disregarding the dilaton without the loss of
generality), the action for the the Kalb-Ramond axion is
\begin{equation}
S_{KR} = \int {\bf H} {^{*}{\bf H}}
\label{4ddual2}
\end{equation}

\noindent The equations of motion are
$d {\bf H} = (\alpha'/16) \Bigl(Tr {\bf R}^2 - Tr {\bf F}^2 \Bigr)$
and $d {^{*} {\bf H}} = 0$.
The first equation is the Bianchi identity, derived from the
definition of the field strength form,
and the second is the Euler-Lagrange equation, obtained from varying
the action. The Euler-Lagrange equation is solved trivially by setting
${\bf H} = {^{*}d}{\bf b}$, and the Bianchi identity translates into a
non-homogeneous Klein-Gordon equation for ${\bf b}$:
$d{^{*}d}{\bf b} = (\alpha'/16)
\Bigl(Tr {\bf R}^2 - Tr {\bf F}^2 \Bigr)~$.
This is typically implemented at the level of the action as follows:
one first takes the vector field ${\bf V} = {^{*}{\bf H}}$, and
recalling that for a p-form in a D-dimensional Minkowski space
${^{**}{\bf \alpha}_p} = -(-1)^{p(D-p)}{\bf \alpha}_p$,
rewrites the Bianchi identity as
$d{^{*}{\bf V}} = (\alpha'/16) \Bigl(Tr {\bf R}^2 - Tr {\bf F}^2 \Bigr)~$.
Then the action is
\begin{equation}
S_{KR} = \int \Bigl\{ -{\bf V} {^{*}{\bf V}} + 2{\bf b}
\Bigl(d{^{*}{\bf V}} - \frac{\alpha'}{16}(Tr {\bf R}^2 - Tr {\bf F}^2 )
\Bigr) \Bigr\}
\label{4ddual4}
\end{equation}

\noindent The (pseudoscalar) field ${\bf b}$ plays role of the Lagrange
multiplier to enforce the Bianchi identity. However, one can then
partially integrate the ${\bf b} d{^{*}{\bf V}}$
term, dropping the (classically
irrelevant) boundary term, and treat ${\bf V}$ as the Lagrange
multiplier. Integrating it out yields the standard pseudoscalar axion
action
\begin{equation}
S_{KR} = \int \Bigl( d{\bf b}{^{*}d{\bf b}}
- \frac{\alpha'}{8}{\bf b}
\bigl(Tr {\bf R}^2 - Tr {\bf F}^2 \bigr) \Bigr)
\label{4ddual5}
\end{equation}

It is not difficult to show that the above method can be
replaced with the following procedure, which is classically completely
equivalent. It can be dubbed the "first order formulation" for the
axion field. Namely, instead of looking at the action (\ref{4ddual2}), one
can look at the formally extended action
\begin{equation}
S_{KR} = \int \Bigl\{ {\bf H}{^{*}{\bf H}} -
2{\bf V} \Bigl( {\bf H} - d{\bf B} - \frac{\alpha'}{16}
({\bf \Omega}_L - {\bf \Omega}_{YM})\Bigr) \Bigr\}
\label{4ddual6}
\end{equation}

\noindent where the definition of ${\bf H}$ as
a field strength associated with
${\bf B}$ is implemented with the help of a 1-form
Lagrange multiplier ${\bf V}$.
In this setting, all three variables ${\bf H}$,
${\bf B}$ and ${\bf V}$ are independent.
The Euler-Lagrange equation for
${\bf B}$ then tells that the "connection" ${\bf V}$
has vanishing curvature, $d{\bf V} = 0$, and hence is a "pure
gauge", ${\bf V} = d{\bf b}$.
Then the field strength ${\bf H}$ can be treated equally well as a Lagrange
multiplier and integrated out, leaving precisely the action (\ref{4ddual5}),
up to a boundary term. Hence the two methods {\it are completely equivalent}
in four and more dimensions.

In three dimensions, the first method actually is meaningless,
because the Bianchi constraint on ${\bf H}$ is satisfied vacuously, being
identically zero for all cases. Therefore the dualization should be
performed along the lines of the second method. As one is
interested here in the theory described by (\ref{act}), the dilaton dependence
will be restored. The axion part of the action (\ref{act}) is
\begin{equation}
S_{KR} = \int e^{-{\bf \Phi}} {\bf H}{^{*}{\bf H}}
\label{3dd1}
\end{equation}

\noindent The Bianchi identity for ${\bf H}$ can now be ignored, because it
contains no information about it. Implementing the definition of ${\bf H}$
as a constraint, along the lines of the procedure described in relation
to (\ref{4ddual6}), gives
\begin{equation}
S_{KR} = \int \Bigl\{ e^{-{\bf \Phi}} {\bf H}{^{*}{\bf H}}
+ 2{\bf Q} \Bigl( {\bf H} - d{\bf B} - \frac{\alpha'}{16}
({\bf \Omega}_L - {\bf \Omega}_{YM})\Bigr) \Bigr\}
\label{3dd2}
\end{equation}

\noindent where ${\bf Q}$ is now a 0-form, since the manifold in question is
three-dimensional. The classical equations of motion associated with the
degrees of freedom ${\bf B}$, ${\bf H}$ and ${\bf Q}$ are, respectively,
\begin{eqnarray}
d{\bf Q}&=&0 \nonumber \\
{^{*}{\bf H}}&=&- e^{{\bf \Phi}} {\bf Q}  \\
{\bf H}&=&d{\bf B} + \frac{\alpha'}{16}
\Bigl({\bf \Omega}_L - {\bf \Omega}_{YM} \Bigr) \nonumber
\label{3dd3}
\end{eqnarray}

\noindent Thus, ${\bf Q}$ is constant (hereafter denoted $Q$),
and $Qd{\bf B}$ is a boundary term,
which can be dropped from (\ref{3dd2}). The field strength can be
integrated out from the action, and the final result is that in the dual form
the axion action is
\begin{equation}
S_{KR} = \int \Bigl( \frac{Q^2}{6} e^{\bf \Phi}
\epsilon_{abc}{\bf e}^a{\bf e}^b{\bf e}^c - \frac{\alpha'}{8}
Q  \bigl( {\bf \Omega}_L - {\bf \Omega}_{YM}\bigr) \Bigr)
\label{3dd4}
\end{equation}

\noindent The original 3D stringy action (\ref{act}) can be
rewritten as
\begin{eqnarray}
S~=~-\frac{1}{2} \int \Bigl\{ e^{-{\bf \Phi}}
&\Bigl(&\epsilon_{abc}{\bf e}^a {\bf R}^{bc} + d{\bf \Phi}~{^{*}d}{\bf \Phi}
+{\alpha' \over 8} Tr {\bf F}~{^{*}{\bf F}} \nonumber \\
&&+ \frac{\Lambda({\bf \Phi})}{3}
\epsilon_{abc}{\bf e}^a{\bf e}^b{\bf e}^c~
- \frac{Q^2}{3} e^{\bf 2\Phi}
\epsilon_{abc}{\bf e}^a{\bf e}^b{\bf e}^c\Bigr)
+ \frac{\alpha'}{4} Q \bigl( {\bf \Omega}_L - {\bf \Omega}_{YM}\bigr)
\Bigr\}
\label{actTMG}
\end{eqnarray}

In general, due to the dilaton dynamics, the action (\ref{actTMG}) does not
have the form of topologically massive gravity coupled to a topologically
massive gauge theory. However, if $\exp{(-{\bf \Phi})} \ne 0$, by a
conformal rescaling one can rewrite it \cite{DY},
so that either the gravitational
or gauge sector appear canonical, but not both,
for the corresponding kinetic terms have different conformal weights. Only
when the total dilaton self-interaction
potential $V(\Phi)$ (all the terms in the action (\ref{actTMG})
except the dilaton kinetic terms) has a minimum and the dilaton decouples,
the action (\ref{actTMG}) resembles the canonical
form of topologically massive gravity coupled to a topologically
massive gauge theory. The resemblance falls short of identity, though,
as the Ricci scalar enters the action (\ref{actTMG}) with the sign
opposite to that of TMG. Therefore, following  \cite{DJT}, one has to
conclude that the massive gravitational excitations are ghost-like,
with the mass $\approx M_P$.

The topological mass terms arise due to the presence of the Chern-Simons
couplings to the axion, and the dynamical triviality of the axion in three
dimensions, where it can only be a topological condensate proportional
to the volume form. The numerical value of the condensate $Q$ is constrained
by requiring that the partition function of (\ref{actTMG}) is gauge invariant.
The inspection of the Chern-Simons sector of the action
then discloses that there occurs a dual quantization of $Q$, if the
theory is to remain gauge invariant under large gauge transformations, too.

The mass terms can be viewed as a product
of the breakdown of the ${\bf B}$-field
gauge invariance. That there is a possiblity to study yet
another symmetry breaking in 3D string theory
can be glimpsed at by looking at the role played by the dilaton. The action
above is independent of the constant background value of the dilaton, which
then can be shifted, $\Phi \rightarrow \Phi + \Phi_0$. Here, $\Phi_0$ can
be understood as the v.e.v. of the quantum dilaton field, and $\Phi$ as
a fluctuation around it. In the limit
when $\Phi_0 = <\Phi> \rightarrow \infty$, which corresponds to the strong
coupling regime of string theory (and represents a consistent classical
solution when there is a cancellation of the divergent
axion contribution $Q^2 \exp(2 \Phi)$, implying
$\Lambda(\Phi) = Q^2 \exp(2\Phi) + v(\Phi)$, with
$\lim_{\Phi_0 \rightarrow \infty} \exp(-\Phi_0) v(\Phi_0) \rightarrow 0$,
as can be easily verified from (4)),
the action (\ref{actTMG}) reduces to the lowest order to
\begin{equation}
S_{KR} = -\frac{\alpha'}{8} Q \int \Bigl( {\bf \Omega}_L
- {\bf \Omega}_{YM} \Bigr)
\label{CS}
\end{equation}

\noindent This action is purely topological, since all the reference to the
metric has disappeared. It looks like a linear combination of two
Chern-Simons gauge theories. There is no reason to distingush
between the contributions any more, since as Witten and Horne and Witten
have argued \cite{WHW},
the sector of conformal gravity in (\ref{CS}), described
by  ${\bf \Omega}_L$, is equivalent on-shell to a
Chern-Simons gauge theory on $SO(3,2)$. Thus, one can think of the theory
defined by (\ref{CS}) as a Chern-Simons gauge theory defined on the group
$SO(3,2) \times G$, where $G$ is the low-energy Yang-Mills group.
This theory then may be assumed to hold consistently
in the unbroken phase of gravity, as pointed out in \cite{WHW}.
Therefore, in this limit, topological gravity becomes merely
a sector of a Chern-Simons gauge theory on a semi-simple Lie group,
a factor of which is the 3D conformal group.

In summary, it was shown that topologically massive gravity as well as a
topologically massive gauge theory can arise from three-dimensional
models containing 2-form fields with anomalous
couplings to gravity and/or gauge fields. One such model, which was
explicitly studied here, is the background field formulation of 3D string
theory to the lowest non-trivial order in the inverse tension expansion.
The inspection of the resulting model has focused on the roles played by
the axion and dilaton fields. While the axion is directly responsible for
the appearance of the topological mass terms, the background value of the
dilaton field plays role of the trigger for the manifest breaking of
conformal invariance. It was argued that, to the lowest order in $\alpha'$,
the strong coupling limit corresponds to the conformally invariant
phase, where the theory can be understood as a Chern-Simons gauge
theory on a semi-simple Lie group. There gravity becomes formally
indistinguishable from the Yang-Mills fields, as it is just a sector
of the gauge group.

\vskip1cm
{\bf Acknowledgements}
\vskip0.5cm
The author would like to thank the Aspen Center for Physics,
where a part of this
research was carried out, for kind hospitality.  Thanks are also due to
B. Campbell for interesting discussions.
This work has been supported in part by
the Natural Science and Engineering Research Council of Canada.

\end{document}